# Developing BCS ideas in the former Soviet Union

## Lev P. Gor'kov

National High Magnetic Field Laboratory, Florida State University, Tallahassee, FL 32310, U.S.A.

The essay is an attempt to re-create the wonderful scientific atmosphere that emerged after the basic BCS ideas first arrived in Russia in 1957. It summarizes the most significant contributions to the microscopic theory of superconductivity by Russian physicists during the next few years that have given the theory its modern form.

## 1. Introduction

These brief historical notes are a revised version of a lecture delivered by the author during the jubilee Symposium "BCS@50" held at University of Illinois at Urbana-Champaign (UIUC) in October, 2007.

I confine myself to a comparatively short period between the end of 1957 and the early 1960s. It was the time when the underlying ideas and methods of the microscopic theory of superconductivity itself were under careful examination in parallel with experimental efforts to test the BCS theory's predictions.[1-3]

The most significant theoretical accomplishments during that early period were the canonical transformation version of BCS (Bogolyubov)  , and the elaboration of superconductivity theory by using Quantum Field Theory (QFT) methods, as well as the identification of the Cooper pair wave function as the symmetry order parameter in the superconducting phase (Gor'kov).  The history of these events in Russia, during the days of the former Soviet Union (USSR), is not known well in the West.

## 2. Science in the USSR

After the end of the Second World War exact sciences in the USSR remained practically isolated from the West. By 1957 direct contacts between scientists were cut to minimum on both sides. Soviet authorities strictly limited the Institutes of the Russian Academy's subscriptions to Western journals, partially in an attempt to save on foreign currency. New issues of Physical Review used to arrive after a delay. Acquiring any modern Western experimental equipment was difficult due to restrictions imposed by the US on the trade with the USSR. While these restrictions had their anticipated negative impact by introducing an unnecessary parallelism in research, ironically, it also strengthened the traditional originality of Russian Science. Russian Science continued to maintain a leading position in many areas of physics, theoretical physics and mathematics.

Participation in atomic projects had boosted the prestige and secured a comparative independence for physicists in Soviet society. Science in the USSR was organized quite differently than in the Western university system. Research, at least most of non-classified research, was not concentrated in universities as in the West, but mainly at the Institutes of the Academy of Sciences of the USSR. Soviet Physics and Mathematics also remained sound and healthy due to concerted governmental effort to attract and educate a whole new generation of young scientists. The pre-war generation of physicists remained active and at the height of their power; those included such figures as Petr(Pyotr) Kapitza, Lev Landau, and Igor Tamm, to name just a few.

Of especial significance for the exact sciences in the USSR was the tradition of Scientific Schools. In connection with the theory of superconductivity I mention only two of them: the Landau School and the group headed by N.N. Bogolyubov.

Among the physicists that formed the Landau School there were many internationally respected scientists, the heads of theoretical groups at the Institutes of the Academy of the USSR. Affiliation with the Landau School had never born an official character. Most common to the Landau School was a broad area of interests. The School united theorists who shared a no-nonsense attitude toward physics, and a high level of the professionalism, including, in particular, an easy use of the mathematical apparatus. The carefully elaborated educational process played an important role in achieving of all this. To be affiliated with the Landau School, one had to pass the exams of the famous Landau "Theoretical Minimum" that comprised a considerable part of all ten volumes of the current Landau and Lifshitz *Course of Theoretical Physics*. As of today, many theorists got through the school of the "Theoretical Minimum" exams; the beginning of Landau's efforts to educate his disciples ran back to his stay in Kharkov in the 1930s.

The only official position Landau had held since 1939 was that of the head of the small theory department at the Kapitza Institute for Physical Problems in Moscow. For a long time his group consisted of two people: E. M. Lifshitz. I. M. Khalatnikov, with A. A. Abrikosov admitted around 1950. Three younger staffers, L. P. Gor'kov, I. E. Dzyaloshinskii and L. P. Pitaevskii, joined the Landau group in the mid-1950s.

The personal style of N. N. Bogolyubov and his interests in Statistical Physics were of a more mathematical character. In the mid-1950s he was the head of the Theoretical Department at the Steklov Mathematical Institute of the Soviet Academy of Sciences in Moscow, while still preserving posts in Kiev, Ukraine. In 1956 the Soviet Government organized the Joint Institute for Nuclear Research (JINR) in Dubna, a city not far from Moscow, and Bogolyubov became its first Director. JINR was designed as an international organization open to scientists from countries belonging to the Soviet Block. The participation of European countries with somewhat more liberal traditions, such as Poland, Checkoslovakia and Hungary, made it possible for scientists at JINR to enjoy a relative amount of freedom in their contacts with the West. Unlike Landau himself and members of his group, scientists from JINR traveled abroad.

In studies of the basics of superfluidity and superconductivity Soviet physicists were at the forefront at that time. The Institute for Physical Problems (The Kapitza Institute) maintained a leadership position in the area of low temperature physics. Superfuidity of He II had been discovered in 1938 by Kapitza and defined as the capability of the liquid to flow along narrow capillaries without viscosity below the *lambda*-point, $T_\lambda = 2.19 K$ . The notion that superfluidity and superconductivity were two tightly related phenomena had become common soon after Landau developed the theory of He II (1940-41).

## 2.1. Superfluidity and Superconductivity

In papers on helium Landau introduced one of the most important paradigms of modern Statistical Physics.[4] The basic concept was that low temperature properties of any macroscopic system could be described in terms of a gas of excitations called quasiparticles (*qps*). These *qps* are brought about as the result of interactions and, generally speaking, do not have much in common with the properties of free particles of which the system is formed. The helium *qps* must obey Bose statistics. Electrons in metals obey Fermi statistics.

Low energy excitations in He II are quantized sound waves, phonons, with a linear dispersion: *E(p)= cp*. Landau proceeded, however, with a general form of the energy spectrum, *E(p),* which at a higher momentum may have an arbitrary shape (to account for experimental data Landau later assumed for liquid He II the spectrum with the so-called "roton" minimum).[4]

Consider $T = 0$ . There are no excitations in the liquid as long as the helium is at rest. Let then helium move along the capillary with a velocity *V*. Forming an excitation inside the moving liquid, $E(p)$, with the momentum *p*, its energy, $E'(p)$, in the reference system where the capillary is at rest, becomes, in the accordance with the Galileo's law:

$$E'(p) = E(p) + (\vec{p} \bullet \vec{V}) \tag{1}$$

In other words, forming an excitation in a moving liquid may cost less in energy (let $(\vec{p} \bullet \vec{V}) = -pV < 0$). The slope of the straight line, $pV$, drawn in the $(E, p)$-plane, increases with the velocity increase, $V$, and the line will finally touch the spectrum curve, $E(p)$, at some $p$ so that $E(p)$- $pV$ =0. The condition determines the very moment when creation of excitations in helium becomes energetically possible for the first time, and at higher $V$ the system would start heating itself (to "dissipate") by producing the *qps* excitations. Hence, the critical velocity, $V_{cr}$, is defined as:

$$V_{cr} = \min(E(p)/p) \tag{2}$$

Although the initial slope, $E(p) \cong cp$, could guarantee superfluidity, in the real He II $V_{cr}$ is less than the speed of sound, $c$, and is determined by the "roton" part of the spectrum.[4]

Bogolyubov contributed to the theory of superfluidity in his famous 1947 paper[5] on the energy spectrum of the non-ideal Bose gas. He showed that in the presence of a weak repulsive interaction between the Bose particles, the parabolic energy spectrum of the ideal gas, $E(p) = p^2/2M$, undergoes a fundamental change. Namely, at low momenta, due to the finite compressibility, the energy spectrum becomes linear in $p$, corresponding to the emergence of the "phonon" (sound) mode. Therefore, in accordance with Landau's arguments above, the non-ideal Bose gas possesses superfluidity. In his 1947 paper[5] Bogolyubov successfully exploited the very concept of the Bose-condensate phenomenon, to wit, the existence of the coherent quantum state with zero momentum occupied by the *macroscopic* number of particles. Therefore the creation and the annihilation operators, $a^+_0$ and $a_0$, for particles with zero momentum can be handled as c-numbers:

$$< N_0 + 1 \mid a_0^+ \mid N_0 > \approx < N_0 - 1 \mid a_0 \mid N_0 > \propto (N_0)^{1/2} \tag{3}$$

($N_0$ is the number of particles in the condensate).

Returning now to superconductivity in metals, the Landau mechanism could then explain the superconductivity phenomenon as the superfluidity of the electronic liquid (abstracting from the magnetic fields produced by charge currents) if the Fermi spectrum, $E(p) = v_F(p - p_F)$, of metallic electrons in the normal phase below $T_c$ were modified by the emergence of a small energy gap at the Fermi level. Experimental indications in favor of such gap[6] had indeed started to accumulate by 1956.

### 3. BCS receives the immediate recognition in the USSR

Before proceeding further, it should be noted that, unlike in the West, the results and ideas of BCS theory[3] were recognized at once in Russia, at least by theorists. This fact deserves a few comments.

First, BCS arrived at the gapped spectrum at $T = 0$[3], as it was expected for the Landau mechanism to work. The retarded character of attraction between electrons *via* the virtual phonon exchange[7] seemed to be helpful for a reduction of the screened Coulomb interaction.[8]

Secondly, the proof by Cooper of the absolute instability of the Fermi Sea in the presence of an arbitrary *weak* electron-electron attraction[1] was perceived as a *qualitative* idea, capable of explaining why superconductivity was wide-spread along the Mendeleev Chart, although the temperatures of transition, $Tc \sim 1$-10K, for the superconductors known at that time were so low compared to the typical energy scales in metals.

There were no "great expectations" as to obtaining an "ideal" agreement between theory and experiment because of the well-known strong anisotropy of the Fermi surfaces in metals (actually, as we know now, the agreement turned out to be remarkably good for the isotropic model leading, for instance, to the non-trivial explanation of such a tiny feature as the Hebel-Slichter peak[9] in 1957).

The Russian community became familiar with the main ideas of BCS in the much simpler language of the Bogolyubov canonical transformation or the formulation of the new theory of superconductivity in terms of Quantum Fields Theory methods developed by Gor'kov. The Soviet theorists, for instance, never had any difficulties with the gauge invariance of the theory.

Finally, the transparency of the theory and its beauty was a very strong argument in its favor, at least in the eyes of Landau and his group.

### 4. Bogolyubov canonical transformation

The first of the Bogolyubov papers[10] was submitted to the Russian Editors on October 10, 1957. Bogolyubov had formulated his method at $T = 0$ by departing from the phonon Frőhlich Hamiltonian:

$$H_{Fr} = \sum_{k,s} E(k) a_{ks}^{+} a_{ks} + \sum_{q} \omega(q) b_{q}^{+} b_{q} + H'$$

$$H' = \sum_{k, q = k' - k, s} g \left\{ \frac{\omega(q)}{2V} \right\}^{1/2} a_{ks}^{+} a_{k's} b_q^{+} + h.c. \tag{4}$$

(We use the units with Plank's constant $\hbar \equiv 1$).

The phonons are then integrated out by going to the second approximation in **g**, so that, actually, Bogolyubov solved exactly the same Hamiltonian as the BCS paper[3].

It[10] was suggested that the operators $a_{ks}^{+}$ and $a_{ks}$ in the superconducting state transform into a set of operators for a new *qps*:

$$a_{k,1/2} = u_k \alpha_{k,0} + v_k \alpha_{k,1}^{+}$$
$$a_{k,-1/2} = u_k \alpha_{-k,1} - v_k \alpha_{-k,0}^{+} \tag{5}$$
$$u_k^2 + v_k^2 = 1$$

The substitution of (5) into the initial Hamiltonian leads to non-diagonal terms, $(\alpha_k \alpha_{-k} + h.c.)$.

To find the coefficients $u_k, v_k$, Bogolyubov resorted to what he called the "principle of compensation of "dangerous diagram". Here is how the "principle" was formulated in his paper.[10] Suppose one considered the perturbation corrections to the new vacuum. Taking separately, non-diagonal terms in the transformed Hamiltonian would produce in the intermediate states the denominators of the form: $2E(k')$. Quoting the Russian text:[10] "the energy denominator $1/2E(k')$ becomes dangerous for integrating"... Then again:[10] "Thus, in the choice of the canonical transformation, it must be kept in mind that it is necessary to guarantee the mutual compensation of the diagrams which lead to virtual creation from the vacuum of pairs of particles with opposite momenta and spins".

It is difficult to understand Bogolyubov's reasons for using this rather obscure wording. By "dangerous" terms with $2E(k')$ in the denominators, Bogolyubov in all probability was alluding to the logarithmic divergences in the Cooper channel. Bogolyubov[10] does not reference the Cooper article[1], though he does reference the BCS letter.[2]

It was soon realized that, rather than utilizing Bogolyubov's reasoning in terms of "dangerous" diagrams, one should merely argue that in the new (superconducting) state at $T = 0$ all terms in the Hamiltonian containing products of operators $\alpha_k \alpha_{-k}$ must give zero when applied to the wave function of the new vacuum.

Calculating various superconducting properties in terms of Bogolyubov *qps* turns out to be much simpler, hence its popularity (the method was soon generalized by many authors, including Bogolyubov himself, to the case of finite temperatures). In Russian literature the theory of superconductivity was often referred to as the "Bardeen, Cooper and Schrieffer and Bogolyubov theory".

The main triumph of BCS, from a fundamental point of view, was its derivation of the gapped energy spectrum from whence superconductivity follows in accordance with the Landau criterion. Note in passing that the BCS theory meant clean superconductors. Neither BCS, nor the Bogolyubov formulation provides the definition of the order parameter and its symmetry in the superconducting state. The theory had yet to be generalized to spatially non-homogeneous problems, especially for alloys. Besides, it was not clear how to go beyond the weak coupling approximation of BCS. This was all achieved within the framework of the Quantum Field Theoretical (QFT) method.[11]

## 5. Quantum Field theory methods ($T = 0$)

Quantum Electrodynamics was still a busy field, even in the early 1950s. It would be untimely to discuss this activity here. For our purposes, suffice to say that many from those days knew the Feynman diagrammatic methods perfectly well. For instance, my PhD thesis was on the "Quantum Electrodynamics of charged particles with zero-spin" in 1956.

Applying the diagrammatic quantum field approach to the needs of condensed matter physics (at $T = 0$) began in Russia around 1956-57. Indeed, the generalization of the methods of Quantum Field Theory to Fermi systems looked rather straightforward with the Fermi sea playing at $T = 0$ the role of the vacuum. A systematic discussion of the diagrammatic rules and some applications has been published in Ref. 12, but the technique actually was in use even before. Thus, the electron-phonons interactions in metals had been studied by Migdal[13] in 1957. Landau had applied the method of microscopic derivation to the theory of the Fermi Liquid[14] in 1958. It is also worth adding that the atmosphere at the Landau School was very friendly, and after talks at the Landau seminar the results were discussed broadly before publication.

## 6. Bogolyubov talk at the Landau Seminar

By the fall 1957 I was the junior scientist in the Landau group and had published papers on Quantum Electrodynamics, Hydrodynamics and helium.

Sometime in October 1957 it got abroad that N. N. Bogolyubov had finished the paper on the theory of superconductivity. He was invited to give a talk at the Landau Seminar in the Kapitza Institute.

The seminar started with somewhat heated debates. Bogolyubov focused on the formal part, i.e., on the details of his method of the canonical transformation, Landau, as usual, preferred to first hear the physics behind it. It was difficult for him to get through the formal Bogolyubov's "principle of compensation of "most dangerous" diagrams". Indeed, as we have seen it above, the "principle" itself was not very transparent, to say the least! Landau wanted to know the nature of the new vacuum. Here I need to explain that neither the Cooper paper[1] published in 1956, nor the short BCS letter[2] had attracted the attention of anyone in the Landau group. After the seminar break, N.N. Bogolyubov finally resorted to mentioning Cooper's result. He repeated the calculations by Cooper on the blackboard. Its transparent physics had the immediate effect of pacifying Landau.

As I was listening, it crossed my mind that the instability of the Fermi sea in the presence of a weak attraction between electrons that results in the spontaneous formation of pairs, also involves the emergence of a bosonic degree of freedom, and I decided to play with the idea.

## 7. Developing the QFT approach for theory of superconductivity

### 7.1. *T=0 and beyond*

I cannot help but show the title and the abstract of my first paper[11] on superconductivity, which as I understand it, did not attract much of attention from main players in the West. The paper came out in the Russian Journal ZhETF in March of 1958. By that time references to BCS could be added.

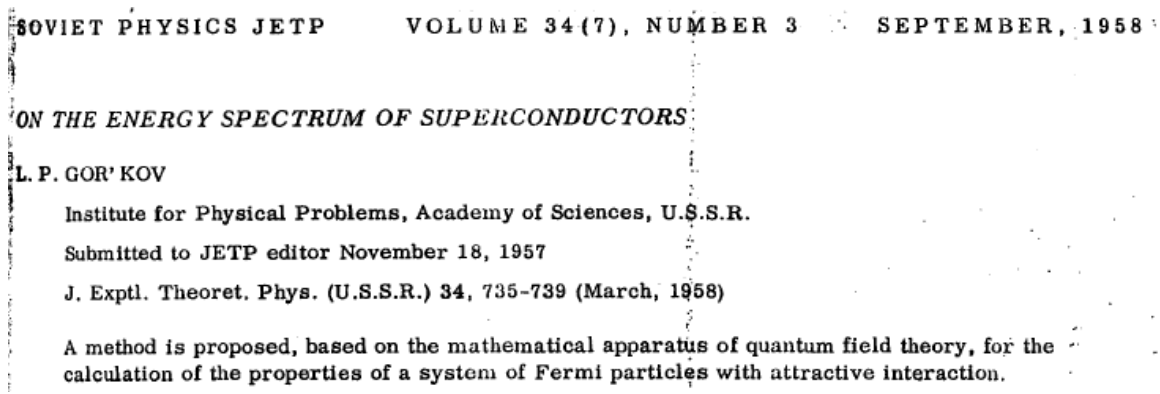

    A method is proposed, based on the mathematical apparatus of quantum field theory, for the calculation of the properties of a system of Fermi particles with attractive interaction.

I had been working with the same four-fermion interaction Hamiltonian as BCS[3] and Bogolyubov:

$$H_{int} = (1/2) \sum_{k,k';\sigma,\sigma'} V_{k,k'} \hat{c}^+_{k,\sigma} \hat{c}^+_{k'\sigma'} \hat{c}_{k',\sigma'} \hat{c}_{k,\sigma} \qquad (6)$$

with $V_{k,k'} = g < 0$ negative and non-zero only for the energies of electrons inside an interval $\varepsilon(k), \varepsilon(k') < \omega_D$, a typical phonon frequency. As far as the energies of all the electrons involved would remain well below $\omega_D$, $(T_c << \omega_D)$, the interaction in the Hamiltonian (6) could be considered as the local one in space:

$$H = \int \left\{ -(\bar{\psi}^+ \frac{\Delta}{2m} \bar{\psi}) + \frac{g}{2} (\bar{\psi}^+ (\bar{\psi}^+ \bar{\psi}) \bar{\psi}) \right\} d^3 r \qquad (7)$$

In the spatial representation the field operators, $\bar{\psi}_\alpha(r), \bar{\psi}^+_\theta(r)$, are:

$$\bar{\psi}_\alpha = V^{-1/2} \sum_{k,\sigma} \hat{c}_{k\sigma} s_{\sigma\alpha} \exp(ikr)$$

$$\bar{\psi}^+_\beta = V^{-1/2} \sum_{k,\sigma} \hat{c}^+_{k\sigma} s^*_{\sigma\alpha} \exp(-ikr) \qquad (8)$$

It was then straightforward to write down the equations for the field operators in the Heisenberg representation:

$$\{i\partial/\partial t + \Delta/2m\} \bar{\psi}(x) - g(\bar{\psi}^+(x)\bar{\psi}(x))\bar{\psi}(x) = 0$$

$$\{i\partial/\partial t - \Delta/2m\} \bar{\psi}^+(x) + g\bar{\psi}^+(x)(\bar{\psi}^+(x)\bar{\psi}(x)) = 0 \qquad (9)$$

(Note the notations $x = (\boldsymbol{r}, t)$).

The Green function is defined as usual:

$$G_{\alpha\beta}(x - x') = -i < T(\bar{\psi}_\alpha(x)\bar{\psi}^+_\beta(x')) > \qquad (10)$$

When the first of Esq. (9) is applied to the operator $\bar{\psi}_\alpha(x)$ in the definition (10) of $G_{\alpha\beta}(x - x')$, one obtains the equation that contains the average of a block of four of the field operators. The main physical idea then was that new terms must appear in this product, those of a Bose condensate type, to account for the bosonic degree of freedom that emerges due to the presence of Cooper pairs at $T = 0$. Correspondingly, the aforementioned block of the four field operators had been decoupled as:

$$\langle T(\tilde{\psi}_\alpha(x_1)\tilde{\psi}_\beta(x_2)\tilde{\psi}_\gamma^+(x_3)\tilde{\psi}_\delta^+(x_4))\rangle = -$$
$$\langle T(\tilde{\psi}_\alpha(x_1)\tilde{\psi}_\gamma^+(x_3))\rangle\langle T(\tilde{\psi}_\beta(x_2)\tilde{\psi}_\delta^+(x_4))\rangle \tag{11}$$
$$+\langle T(\tilde{\psi}_\alpha(x_1)\tilde{\psi}_\delta^+(x_4))\rangle\langle T(\tilde{\psi}_\beta(x_2)\tilde{\psi}_\gamma^+(x_4))\rangle$$
$$+\langle N\,|\,T(\tilde{\psi}_\alpha(x_1)\tilde{\psi}_\beta(x_2))\,|\,N+2\rangle\ \langle N+2\,|\,T(\tilde{\psi}_\gamma^+(x_3)\tilde{\psi}_\delta^+(x_4))\,|\,N\rangle$$

Contributions from the first two terms into the equation of the Green function could be omitted assuming the weak coupling limit, but the last two averages in Eq.(11) introduced into the theory the two *anomalous* functions that are non-zero *only* in the superconducting state:

$$\langle N\,|\,T(\tilde{\psi}_\alpha(x)\tilde{\psi}_\beta(x'))\,|\,N+2\rangle = \exp(-2i\mu t)F_{\alpha\beta}(x-x')$$
$$\langle N+2\,|\,T(\tilde{\psi}_\alpha^+(x)\tilde{\psi}_\beta^+(x'))\,|\,N\rangle = \exp(2i\mu t)F_{\alpha\beta}^+(x-x') \tag{12}$$

The two functions are akin to the c-numbers, Eq.(3), introduced by Bogolyubov for the Bose gas in his 1947 paper and bear the coherent macroscopic origin. The product of the two last terms in (11) would be proportional to the condensed Cooper pairs' density.

Exponential factors containing the Josephson time dependence in Eq.(12) can be removed by using the chemical potential instead of the total number of particles as the thermodynamic variable: $H \Rightarrow H - \mu N$. In the new variables the system of coupled equations has the following form:

$$\{i\partial/\partial t + \Delta/2m + \mu\}\hat{G}(x-x') - ig\hat{F}^+(0+)\hat{F}^+(x-x') = \delta(x-x') \tag{13}$$
$$\{i\partial/\partial t - \Delta/2m - \mu\}\hat{F}^+(x-x') + ig\hat{F}^+(0+)\hat{G}(x-x') = 0$$

In Eq.(13) we denoted, for instance:

$$\hat{F}_{\alpha\beta}(0+) = \langle \tilde{\psi}_\alpha(x)\tilde{\psi}_\beta(x)\rangle \tag{14}$$

It is easy to see that

$$\hat{F}_{\alpha\beta}(0+) = i(\sigma_y)_{\alpha\beta}F$$
$$\hat{F}^+_{\alpha\beta}(0+) = -i(\sigma_y)_{\alpha\beta}F* \tag{15}$$

The antisymmetric spinor structure of the pair wave function corresponds to the *singlet* pairing. Introduce the notation:

$$\Delta = gF \tag{16}$$

By substituting $i\partial / \partial t \Rightarrow E$ in Eqs.(13) re-written in the momentum representation, the BCS gapped energy spectrum for a homogeneous superconductor will immediately follow as the eigenvaules of the L.H.S. operator for the system (13):

$$E(p) = \pm \sqrt{\left[ v_F^2 (p - p_F)^2 + |\Delta|^2 \right]} \tag{17}$$

One may now summarize some main results obtained in the paper[11] by the above formalism. To begin with, while $|\Delta|$ is the magnitude of the energy gap, *the true order parameter* in the superconducting state is $\Delta$, (or $\Delta^+$), i.e., *the wave function of the Cooper pair* or, more broadly, the anomalous functions themselves. The symmetry broken at the transition is the gauge symmetry, *U(1)*. Eqs.(13) obviously possess the gradient-invariant form when a magnetic field is introduced by the usual substitution: $-i\partial \Rightarrow \left( -i\partial - (e/c)A(r) \right)$.

Interactions with the magnetic field and any other perturbations if added into the Hamiltonian (7) at $T = 0$ can be studied with the routine diagrammatic technique for the matrix composed of the Green function, $G$, and the anomalous functions $F$ and $F^+$.

Expressions for the Free Energy and thermodynamics have been obtained in few lines.[11]

Note that Eq. (16) defines the gap, $\Delta$, self-consistently through the $F$-function at equal arguments. In turn, the expression for the latter in the momentum space one obtains by solving the system of Eqs. (13). There is no need in the variational procedure or in the Bogolyubov "principle of compensation of "dangerous" diagrams".

The Green functions at *finite* temperatures can be unambiguously found from Esq. (13) by making use of the relations between the advanced and retarded (casual) Green functions. For normal metals it was derived by Landau in the form[15]:

$$\operatorname{Re} G(\omega) = -\frac{1}{\pi} \int_{-\infty}^{+\infty} \coth \frac{x}{2T} \frac{\operatorname{Im} G(x)}{\omega - x} dx \tag{18}$$

The provisions imposed by Eq. (18) may complicate calculations at finite temperatures since the automatism of the $T = 0$ diagrammatic technique is now lost. I do not stay on this any longer, because in such cases it is preferable to apply the thermodynamic technique that was soon elaborated for the needs of the superconductivity theory, as described below.

## 7. 2. *Superconducting alloys*

In 1958 A. A. Abrikosov joined me. Together we started the application of the above diagram method to superconducting alloys. We published two papers on Electrodynamics and Thermodynamics of alloys[16, 17].

At calculating the transport and other kinetic properties in normal metals in the presence of defects one routinely uses the well-known Boltzmann Equation. Now, to account for the role of defects in the superconducting state, we faced the necessity of somehow including scattering on defects into the general diagrammatic approach. Without entering into details, first we had to develop[16] the so-called "cross-technique," giving the means to treat scattering of the electrons on defects diagrammatically. Our then newly created[17] "Matsubara" technique[18] (see in Sec.7. 3 below) was applied for the first time to alloys at nonzero temperatures.

I will skip the results that account for the role of defects in the electromagnetic properties of a superconductor (i.e., the Meissner effect). Our calculations showed, in particular, that all Green functions, including the *anomalous* function, $F$ and $F^+$ being averaged over the impurities' positions, acquire in the coordinate representation the exponentially decaying factors describing the loss of the coherence due to the scattering:

$$F(t - t', R) \Rightarrow F(t - t', R) \exp(-R / l) \tag{19}$$

(And similarly for the rest of the Green functions; $l$ is the mean free path). Hence, with the gap being defined, according to Eq. (16), at $R=0$, the thermodynamics of a superconductor does not change in the isotropic model.

This result, first obtained by Abrikosov and myself,[16] is known in the West as "the Anderson Theorem."[19]

## 7. 3. *Developing Thermodynamic QFT Technique for Statistical Physics*

Although the transition temperature, $T_c$ of most early superconductors was rather low, the temperature dependence of different superconducting characteristics had been extensively studied experimentally. The needs of the theory of superconductivity had propelled in 1958 the broad implementation of diagrammatic methods for the general Statistical Physics.[18]

At first, in 1955 Matsubara[20] showed that there is a formal analogy between the so-called $\hat{S}$ - matrix in the Quantum Field Theory (at $T = 0$) and the expression of the statistical matrix for the Gibbs distribution. The latter can be written in the form:

$$\exp\left[\frac{\mu\hat{N} - \hat{H}}{T}\right] = \exp\left[\frac{\mu\hat{N} - \hat{H}_0}{T}\right] \times \hat{S}(1/T) \tag{20}$$

The "S-matrix" here is:

$$\hat{S}(1/T) = \hat{T}_\tau \exp(-\int_0^{1/T} \hat{H}_{int}(\tau)d\tau) \tag{21}$$

The inverse temperature, $1/T$, may be taken formally as an imaginary time, $\tau$. The diagrammatic expansion can now be developed for the new "Green functions" defined as:

$$\overline{G}(1,2) = -\frac{\langle\langle\hat{T}(\widehat{\overline{\psi}}(1)\widehat{\psi}^+(2)\hat{S})\rangle\rangle}{\langle\langle\hat{S}\rangle\rangle} \tag{22}$$

In Eq. (22) the double brackets $\langle\langle...\rangle\rangle$ mean the *grand canonical ensemble* average; the field operators now depend on the space coordinates and on the *imaginary* "time."[20]

The difficulty is that in Eq. (21) the "imaginary time" $\tau$ varies only inside the finite interval ($0 < \tau < 1/T$) and the representation in the form of the Fourier integrals for the "time variables" is not possible. Abrikosov et al.[18] suggested instead to expand the temperature Green function (22) into the *Fourier Series*:

$$\overline{G}(\tau - \tau') = T\sum_n \overline{G}(\omega_n)\exp(-i\omega_n(\tau - \tau')) \tag{23}$$

where $\omega_n = \pi T n$ with n-even for Bose system, n-odd for the Fermi case

The diagrammatic rules for diagrams with the new Green functions, $\overline{G}(\omega_n)$, turn out to be basically the same as at $T = 0$, except differences in the numeric coefficients and the fact that in all matrix elements for diagrams the *summations* now substitute for the integrations over "frequency" variables. The analytical continuation in the complex frequency plane, $z = \omega$, allows us to connect the thermodynamic Green functions (23) calculated *at the points*, $z = i\omega_n$, along the *"imaginary"* (or the Matsubara) axis, with the casual Green functions on the real axis.[18]

Gor'kov equations (13) in the coordinate space can now be written in the thermodynamic notation:[21]

$$\left\{ i\omega_n + \frac{1}{2m}(\frac{\partial}{\partial r} - i\frac{e}{c}A(r))^2 - \mu \right\} \overline{G}(\omega_n; r, r') + \Delta(r)\overline{F}^+(\omega_n; r, r') = \delta(r - r')$$

$$(24)$$

$$\left\{ -i\omega_n + \frac{1}{2m}(\frac{\partial}{\partial r} + i\frac{e}{c}A(r))^2 - \mu \right\} \overline{F}^+(\omega_n; r, r') - \Delta^*(r)\overline{G}(\omega_n; r, r') = 0$$

Here the "gaps" are determined through the anomalous functions with the coinciding arguments as, for instance:

$$\Delta(r) = gT\sum_n \overline{F}(\omega_n; r, r) \qquad (25)$$

With the gauge transformation $A(r) \Rightarrow A(r) + \nabla\phi(r)$ the "gap" transforms correspondingly: $\Delta(r) \Rightarrow \Delta(r)\exp(i2\phi(r))$.

We have mentioned that the use of the Bogolyubov *qps* significantly simplifies derivation of the Free Energy in the superconducting phase compared to.[3] However, it is also not so easy to find the *qps* energy spectrum in the field's presence. Another example is given by alloys where defects are randomly distributed along the superconducting sample. In these instances the following expression for the Gibbs' potential is extremely helpful:[21]

$$\Omega_S - \Omega_N = -\int d^3 r \int_0^g \frac{\delta g}{g^2} \mid \Delta(g; r) \mid^2 \qquad (26)$$

## 7.4. *Ginsburg-Landau Equations from Microscopic Theory*

The Ginsburg-Landau theory[22] was one of the most important phenomenological breakthroughs prior to the creation of the microscopic theory of superconductivity. The theory[22] made it possible to account for countless data on non-linear magnetic properties of superconductors in the good agreement with the main experimental findings. It was therefore important to find out whether it can be substantiated on the microscopic level. This had been done early in 1959 in two of my papers,[21, 23] for clean superconductors and for superconducting alloys, correspondingly. Without entering into the actual calculations, I only list below the main results.

Close to the temperature of transition, $|T - T_c| \ll 1$, Eq. (25) can be simplified. Indeed, near $T_c$ $\Delta(r)$ is small and Eqs. (24) for $\overline{G}$ and $\overline{F}, \overline{F}^+$ can be solved perturbatively. When calculating $\overline{F}(\omega_n; r, r)$ in Eq.(25), one can restrict oneself by a few non-zero terms in $\Delta(r)$, $\Delta^*(r)$ and $A(r)$.[21] For clean superconductors such expansion results in the equation of the form:

$$\left\{ \frac{1}{2m}(\partial - i\frac{e^*}{c}A(r))^2 + \frac{1}{\lambda}\left[ \frac{T_c - T}{T_c} - \frac{2}{N} |\Psi(r)|^2 \right] \right\}\Psi(r) = 0 \qquad (27)$$

Eq. (27) has the familiar form of the GL-equation[22] for the "gap" parameter:

$$\Psi(r) = \Delta(r)\sqrt{7\varsigma(3)N} / 4\pi T_c.$$

Note e*=2e that stands for the charge of the Cooper pair! ($N$ is the density of electrons, $\lambda = 7\varsigma(3)E_F / 12(\pi T_c)^2$).

The expression for the electrical current is derived in the same way:

$$j(r) = -\frac{ie^*}{2m}\left( \Psi^* \frac{\partial \Psi}{\partial r} - \Psi \frac{\partial \Psi^*}{\partial r} \right) - \frac{e^{*2}}{mc} A |\Psi|^2 \qquad (28)$$

The GL-theory, as it is known,[22] contains one important dimensionless parameter, $\kappa$. In terms of the observable characteristics of a material, Ginsburg and Landau[22] expressed it in the following way:

$$\kappa = (\sqrt{2}e^* / \hbar c)H_{cr}\delta_L^2 \qquad (29)$$

(Here $H_{cr}$ is the thermodynamic critical field and $\delta_L$ is the penetration depth). Now $\kappa$ can be written through the microscopic parameters of a metal:[21]

$$\kappa = 3T_c(\pi / v_F)^{3/2}(c / ep_F)(2 / 7\varsigma(3))^{1/2} \qquad (30)$$

The parameter $\kappa$ determines the behavior of a superconductor in strong magnetic fields. Abrikosov showed[24] that depending on whether $\kappa$ is small or large, a superconductor would belong to the one of two classes: Type I or Type II, correspondingly. From Eq. (30) one may conclude that the microscopic theory imposes no limitations on the $\kappa$-value, so that even a pure metal can be a Type II superconductor. I predicted[21] that this could be the case for the *elemental* niobium, Nb.

Type II superconductivity is more common in alloys. The GL equations for an alloy, derived in,[23] look quite similar to (27, 28) but with the coefficients now depending on the transport meantime, $\tau_{tr}$ :

$$\left\{ \frac{1}{2m}(\partial - i\frac{e^*}{c}A(r))^2 + \frac{1}{\lambda_\tau}\left[\frac{T_c - T}{T_c} - \frac{2}{N\chi(\rho)} \mid \Psi(r)\mid^2 \right]\right\}\Psi(r) = 0 \quad (31)$$

$$j(r) = -\frac{ie^*}{2m}\left(\Psi^*\frac{\partial\Psi}{\partial r} - \Psi\frac{\partial\Psi^*}{\partial r}\right) - \frac{e^{*2}}{mc}A\mid\Psi\mid^2$$

The GL wave function is now connected with the "gap" as:

$$\Psi(r) = \left[\chi(\rho)7\varsigma(3)N / 16\pi^2 T_c^2 \right]^{1/2}\Delta(r) \quad (32)$$

and:

$$\chi(\rho) = \frac{8}{7\varsigma(3)\rho}\left[\frac{\pi^2}{8} + \frac{1}{2\rho}\left\{\psi(\frac{1}{2}) - \psi(\frac{1}{2} + \rho)\right\}\right] \quad (33)$$

In Eq. (33) $\rho = 1 / 2\pi T_c\tau_{tr}$ , and $\psi(z)$ is the logarithmic derivative of the $\Gamma$-function.

The penetration depth and the value of the GL parameter $\kappa$ also suffer the change:

$$\delta_L = \delta_{L0} / \sqrt{\chi(\rho)} \; ; \; \kappa = \kappa_0 / \chi(\rho) \quad (34)$$

In the short mean free path limit (so-called "dirty" alloys) $\kappa$ is expressible in terms of the observable quantities for the *normal* phase only:

$$\kappa = 0.065ec\gamma^{1/2} / \sigma k_B \quad (35)$$

(Here $\gamma$ and $\sigma$ stand for the coefficient in the linear electronic specific heat and for conductivity, respectively).

Together with the GL and Abrikosov phenomenology[24] the above results are known as the GLAG- theory (after Ginsburg, Landau, Abrikosov and Gor'kov).

### 7.5. *Paramagnetic Impurities*

Alloying by ordinary impurities or defects does not change either the energy gap, or $T_c$, as we have shown in Sec.8.2. In 1960 Abrikosov and Gor'kov[25] studied alloys containing paramagnetic centers, i.e., atoms or ions with the non-zero spin. It turned out that the presence of such centers is detrimental to superconductivity. The transition temperature, $T_c(x)$ decreases with increase of the concentration $x$ of the paramagnetic impurities and vanishes at some critical concentration, $x_{cr}$. This behavior might have been expected, for the potential for scattering on such a defect does now include, in addition, the interaction with the external spin on a center, $\vec{s}$, $V(\vec{p} - \vec{p}') \Rightarrow V(\vec{p} - \vec{p}') + \tilde{V}(\vec{p} - \vec{p}')(\vec{\sigma} \bullet \vec{s})$, that tends to misalign spins of the Cooper pair.

The dependence of $T_c(x)$ on the concentration is given[25] by the equation:

$$\ln(T_{c0}/T) = \psi(\frac{1}{2} + \frac{1}{T_c \tau_s}) - \psi(\frac{1}{2}) \qquad (36)$$

where $1/\tau_s$ is the inverse mean time for scattering with the overturn of the electronic spin and, as in Sec. 7.4., $\psi(x) = \Gamma'(x)/\Gamma(x)$. The value of the critical concentration is given by:

$$\frac{1}{\tau_{c,cr}} = \frac{\pi T_{c0}}{2\gamma} \equiv \frac{\Delta_0}{2} \qquad (37)$$

The gap in density of states (DOS) also decreases with the $x$ increase. It was, however, somewhat unexpected to find that the energy gap in DOS closes *before* the superconductivity is fully destroyed at $x_{cr}$ given by Eq.(37): there exists a narrow range of concentrations below $x_{cr}$ in which superconductivity persists in spite of the zero energy gap in DOS. The result is another proof that it is the non-zero superconductivity *order parameter*, $\Delta$, that controls the capability of a superconductor to carry persistent currents, not the finite gap value. The Landau criterion of Eq. (2) is, hence, of the lesser generality.

## 8. Applications of Quantum Field Theory methods

While for many properties (kinetic properties, such as the sound absorption or thermal conductivity, or the effects, like tunneling, and others) working in terms of the Bogolyubov *qps* was more straightforward, it is not so as far as the electrodynamics of superconductors is concerned, because the new *qps* do not possess the fixed electric charge. There were other fundamental questions for which finding the solution without the diagrammatic methods would be next to impossible.

Thus, the extension of BCS theory to the case of strong electron-phonon coupling had been done by Eliashberg,[26] who in 1960 extended Migdal's paper[13] on electron-phonon interactions for normal metal by introducing, instead of the mere gap parameter, $\Delta$, $\Delta^{+}$, in Eqs. (24) for the "local" Hamiltonian (7), the new self energy parts build up on the anomalous functions $F$, $F+$ :

$$\Delta(\omega_n; p) = T \sum_m \int D(\omega_n - \omega_m; p - p') F(\omega_m; p') \frac{d^3 p'}{(2\pi)^3} \qquad (38)$$

In Eq. (38) $D(\omega_n; k)$ stands for the Green function of phonons:

$$D(i\omega_m; p - p') = g^2 \frac{\omega_0^2(p - p')}{(i\omega_m)^2 - \omega_0^2(p - p')} \ . \qquad (39)$$

As was first demonstrated in[13], the adiabatic approximation, $\omega_0 << E_F$, makes it possible to neglect the so-called "vertex" corrections both in the normal and the anomalous self-energy parts. In the new equations, replacing Eqs. (24), the "gap" functions $\Delta(\omega_n; p)$ and $\Delta^*(\omega_n; p)$ appear together with the self-consistency condition of Eq. (38). The Bogolyubov principle of "dangerous diagrams" could not be relevant in the first place, because there are no divergences in (38) due to the frequency dependence of the phonon Green function.

It is a point to emphasize here that the Quantum Field Theory formulation also makes it easy to generalize to the case of multi-band superconductors, or to take into account the anisotropy of the electronic spectrum.[27] The latter allows the rigorous microscopic derivation of the anisotropic GL equations.[28]

The Gorkov method, as shown by Eilenberger[29] in 1968, can be advanced even further. These considerable simplifications arise with the explicit use of the so-called "quasi-classic" approximation: $T_c, \omega_0 << E_F$.[29] Under this provision that obviously has a quite general character, one may re-write the Gor'kov equations into the master equations for the Green functions integrated over the energy variable:

$$\int \hat{\overline{G}}(\omega_n; \vec{p}, \xi) d\xi = \begin{pmatrix} g(\omega_n; \vec{p}) & f(\omega_n; \vec{p}) \\ f^+(\omega_n; \vec{p}) & \overline{g}(-\omega_n; \vec{p}) \end{pmatrix} \tag{40}$$

(Here momentum $\vec{p}$ may run on the Fermi surface).

The emerging set of non-linear equations is local in space and easier for the numerical analysis and for the treatment of the inhomogeneous problem. We shall not dwell here upon these results.

## 9. Conclusion

The purpose of the sketchy review above was to trace how the development of powerful new theoretical methods made it possible to extend BCS to a much broader class of phenomena. Many new notions and results of a fundamental character came to light since early 1960s, both experimentally and theoretically. The BCS model grew into the mighty theory of superconductivity, one of the most accomplished theories in condensed matter physics. I would like to stress again that, to a considerable extent, the main advances came about through the formulation of the theory in terms of Green functions.

## Acknowledgments

The work was supported by NHMFL through the NSF Cooperative agreement No. DMR-0084173 and the State of Florida.